\documentstyle[preprint,aps]{revtex}
\tighten
\begin{document}
\draft
\preprint{ULLFT--1/94  (March 1994)}

\title
{\bf Dynamical Generation of Mass}

\author{Vicente Delgado}

\address
{Departamento de F\'{\i}sica Fundamental y Experimental,\\
Universidad de La Laguna, E-38203 La Laguna, Tenerife,
Spain.\\ E-mail: vdelgado@ull.es}

\maketitle

\begin{abstract}
An alternative for the Higgs mechanism is proposed.
It predicts the appearance in the broken phase of a
scalar background field which may be interpreted as
describing an almost uniform (i.e., homogeneous and
isotropic) superfluid condensate of decoupled Higgs
bosons. Quantum fields acquire mass as a consequence
of nonperturbative interactions with those particles
condensed in the zero-momentum state (which constitutes
the physical vacuum of the theory) giving rise in turn
to the appearance of density fluctuations. This
mechanism has therefore remarkable cosmological
implications.
\end{abstract}

\narrowtext

Spontaneous symmetry breaking is one of the most widely
observed phenomena in nature \cite{Ander1}. It occurs
when the ground state of a system corresponding to a
particular solution of the equations of motion exhibits
a lower symmetry than the Lagrangian density.

At low temperatures, when thermal fluctuations become
irrelevant, most physical systems undergo phase
transitions toward stable configurations which do not
exhibit the full underlying symmetry of the equations
of motion that govern their dynamical evolution.
Quantum fluctuations are therefore expected to play
an important role in the origin of the instability
of the symmetric phase.
In particular it is well known that the macroscopic
occupation that takes place in quantum condensation
phenomena may be interpreted as a process of symmetry
breaking \cite{ocup1,ocup2,ocup3,ocup4}. The acumulation
of particles in the ground state induces a nonvanishing
vacuum expectation value (VEV) for the corresponding
scalar field, which then acts as the order parameter
required for a complete characterization of the broken
phase.

{}From the perspective of Theoretical Physics spontaneous
symmetry breaking turns out to be of particular interest
because it can provide, in principle, a conceptually
simple mechanism for explaining the complexity of
nature starting from a highly symmetric initial state.
Not surprinsingly, this phenomenon is
one of the basic underlying ideas of both Unified
Gauge Theories and Inflationary Cosmology. In fact,
the only renormalizable gauge theories with massive
vector bosons are gauge theories with spontaneous
symmetry breaking \cite{tHooft}.

In modern Unified Theories \cite{GWS} massive intermediate
bosons are properly introduced by breaking the symmetry by
means of the Higgs mechanism \cite {Higgs,Farhi}.
It basically consists in the introduction of a complex
scalar field $\phi(x)$ subject to an effective potential
of the form
\begin{equation}
V(\phi)=\mu^2\phi^+\phi + \lambda(\phi^+\phi)^2
\end{equation}
with $\mu^2 < 0$. The necessity of introducing a
negative mass squared is the price one has to pay in
order to generate a nonzero stable configuration
$\langle\phi \rangle_0$.

In the present paper it is proposed an alternative for
the Higgs mechanism where spontaneous symmetry breaking
has its origin in the dynamics of the scalar field $\phi$
in the presence of gauge (and matter) fields.
The symmetry breaks once a particular solution of the
equations of motion is chosen among a (one-parameter)
family of possible solutions.

Even though it turns out to be more natural in the
presence of fermion fields \cite{nota}, it is convenient
to consider an abelian U(1) gauge theory in order to
illustrate the mechanism in its simplest context.
This theory, apart from reflecting more clearly the
physics involved, has the additional interest that
in this case the present mechanism basically represents
a covariant generalization of the London theory of
superconductivity. And this analogy with superconductivity
proves to be very useful in order to gain valuable
insights about the nature of the process of symmetry
breaking.
Furthermore, it turns out that the low-energy effective
Lagragian density governing the decoupled scalar sector
takes the same form as in the case of a
SU(3)$\times$SU(2)$\times$U(1) gauge theory \cite{VDB1},
so that most of the cosmological consequences
(directly related to that sector)
may already be extracted by studying this simple case.
For these reasons we shall focus in what follows on a
U(1) gauge theory. Specific application to the
Standard Model will be considered elsewhere
\cite{VDB1,VDB2}.

The U(1) gauge-invariant Lagrangian density
for a complex scalar field $\phi(x)$ reads \cite{libro}
\begin{equation}\label{Lg}
{\cal L}=(D_\mu{\phi})^+({D^\mu{\phi}}) - \mu^2 \phi^+
\phi - \frac{1}{4} {\rm F}_{\mu\nu} {\rm F}^{\mu\nu}
\end{equation}
where now $\mu^2 > 0$ is the mass of the scalar field,
${\rm F}_{\mu\nu} = \partial_\mu B_\nu -
\partial_\nu B_\mu$
is the U(1) field strength tensor, and the
gauge-covariant derivative is given by
$D_\mu=\partial_\mu-igB_\mu$.

It should be emphasized that even though it is possible
to break the symmetry starting with a Lagrangian density
containing a quartic scalar self-interaction,
$\lambda (\phi^+ \phi)^2$, we explicitly take
$\lambda \equiv 0$. The motivation for such a choice lies
mainly in the fact that taking the limit $\lambda
\rightarrow 0$ clearly leads to physical transparency.
Since a $\phi^4$ term is necessary in order to guarantee
perturbative renormalizability in the scalar sector of
the theory, this implies that we choose to work with a
nonrenormalizable scalar sector. Renormalizability has
to be demanded only in the case of a fundamental theory
which must remain valid at arbitrary high energies, and
there exists no indication that the scalar sector of the
Standard Model must satisfy such a requirement. In fact,
superconductivity provides indications just in the
opposite direction. We shall therefore assume that at
low temperatures the scalar field $\phi(x)$ provides an
adequate description of a more fundamental physical
structure, in such a way that the Lagrangian density
($\ref{Lg}$) remains valid up to temperatures of order
$T_c$. Since we are interested in physics in the
broken-symmetry phase, at $T \simeq 0$, in this paper
we shall not concern ourselves about the fundamental
nature of $\mu$ excitations. In fact, as we shall see,
according to the present mechanism the process of mass
generation turns out to be a low-temperature collective
phenomenon where mass appears as a physical parameter
reflecting the cumulative effect of nonperturbative
interactions with those scalar particles macroscopically
condensed in the zero-momentum state. Thus, one expects the
structure of scalar excitations to be irrelevant for the
process of generation of mass. Our physical problem
therefore will reduce to give mass to the vector bosons
preserving (hidden) gauge invariance.

In order to exploit the symmetry of the problem it turns
out to be convenient parametrizing $\phi(x)$ in polar form
\begin{equation}\label{phipar}
\phi(x)= {1 \over {\sqrt 2}} \rho(x) \exp i\xi(x)
\end{equation}
where $\rho(x)$ and $\xi(x)$ are real scalar fields.
In the unitary gauge, which clearly reflects
the particle content of the theory, we have
$\phi(x) \rightarrow \rho(x)/\sqrt{2}$, and the
Lagrangian density can be written
%
\begin{equation}
{\cal L}=\frac{1}{2}(\partial_\mu\rho)(\partial^\mu\rho)+
\frac{1}{2}g^2 A_\mu A^\mu\rho^2-\frac{1}{2}{\mu^2}\rho^2
- \frac{1}{4} {\rm F}_{\mu\nu} {\rm F}^{\mu\nu}
\end{equation}
with
$A_\mu(x)= B_\mu(x) - (1/g)\partial_\mu \xi(x)$.

The Euler-Lagrange equations of motion governing the
coupled dynamics of the fields, are
\begin{eqnarray}
\label{ero}
&&\partial_\mu \partial^\mu \rho(x)= - \mu^2 \rho(x) +
g^2 A^2(x) \rho(x)
\\
\label{efaa}
&&\partial_\alpha \partial^\alpha A^\nu -\partial^\nu
(\partial_\beta A^\beta) = -g^2 A^\nu \rho^2
\end{eqnarray}
where the source of the gauge field,
$\rho^2(x) A^\mu(x)$,
is a conserved current
\begin{equation}
\partial_\mu (\rho^2 A^\mu) = 0
\end{equation}
In what follows we will show that the equations of
motion (\ref{ero})--(\ref{efaa}) admit a family of
solutions which verify
\begin{equation}\label{sol}
A^\mu(x)= {J^\mu(x) \over \rho^2(x)}
\end{equation}
where $J^\mu(x)$ is a conserved current independent of
$\rho(x)$, and a subset of these solutions leads to a
theory with broken symmetry and massive gauge bosons.

Note that
Eq.(\ref{sol}) is nothing but a covariant generalization
of the well-known London formula \cite{London1}, relating
the vector potential with the supercurrents.
The London formula, which can be obtained from the BCS
theory in the long wavelength limit \cite{Super}, plays a
fundamental role in superconductivity, and is essentially
a consequence of the {\em rigidity} of the electronic
superfluid with respect to perturbations \cite{London2}.
It has the importance that it leads to the Meissner
effect, which, as first noted by Anderson \cite{Ander2},
basically reflects the fact that because of nonperturbative
interactions with the medium the gauge quanta dynamically
acquire mass in a superconductor.

As we shall see, the solutions (\ref{sol}) also lead,
in particular, to a rigid (i.e., unperturbed) scalar
field $\rho(x)$, which in the present context must be
interpreted as a decoupled background field.
In fact, the contribution of $A^\mu(x)$ to the dynamical
evolution of $\rho(x)$ becomes suppressed
by inverse powers of $\langle \rho \rangle_0$, so that,
in the weak coupling phase, corresponding to
$\langle \rho \rangle_0 \rightarrow \infty$, the scalar
field becomes virtually unaffected by the presence of
the gauge field, and plays the role of a background
field.
Therefore, the solutions we are interested in actually
describe a nonperturbative broken phase with massive
gauge quanta in a background scalar field.
This fact, in turn, helps us to understand the physics
involved in the process.
Indeed,
Eq.(\ref{sol}), basically showing that in the
nonperturbative phase $A^\mu(x)$ develops a dependence
of the form $1/{\rho^2}$, can be better understood if
one takes into account that eventually $\rho(x)$ behaves
as a classical background field (consider in particular
a cosmological scenario).

By substituting Eq.(\ref{sol}) into Eq.(\ref{ero})
we obtain
\begin{equation}\label{rove}
\partial_\alpha \partial^\alpha \rho(x) =-{\partial \over
\partial \rho}
V_{\rm eff}(x)
\end{equation}
where
\begin{equation}\label{xw1}
V_{\rm eff}(x)=\frac{1}{2} \mu^2 \rho^2(x) + g^2
{J^2(x)\over 2\rho^2(x)}
\end{equation}

As we shall see, the fact that the potential energy density
$V_{\rm eff}(\rho ; J^2)$ depends on the conserved current
only through the composite scalar operator $J^2(x)$, turns
out to be of particular importance in our treatment.

In field theories the vacuum $\vert 0 \rangle$ is defined as
the ground state of the theory. The property of translational
invariance that this state must possess requires the vacuum
expectation values of physical quantities to be constants
independent of space-time coordinates. Therefore, from
Eq.(\ref{rove}) one finds that the vacuum $\vert 0 \rangle$
must verify
\begin{equation}\label{xw2}
\langle 0 \vert {\partial \over \partial \rho} V_{\rm eff}
\vert 0 \rangle = 0
\end{equation}
in agreement with physical intuition. Making use of the
independence of $J^\mu(x)$ on $\rho(x)$, this condition
reads
\begin{equation}\label{evac}
\mu^2 \langle \rho \rangle_0 - g^2 {\langle J^2 \rangle_0
\over \langle \rho^3 {\rangle_0}} = 0
\end{equation}
Note that gauge (and matter \cite{VDB1}) fields make a
nonperturbative contribution in the selection of the
ground state $\vert 0 \rangle$.
Eq.(\ref{evac}) also shows that the properties of the
vacuum depend on quantum fluctuations of physical
currents, the quantum nature of the system playing
therefore an essential role in the process.

Different values of $\langle J^2 \rangle_0$ characterize
different possible solutions and hence different
physical vacua. And due to the fact that $\langle J^2
\rangle_0$ is a constant it is not possible to smoothly
pass from one to another solution, so that we have a
one-parameter family of stable solutions, and the U(1)
symmetry breaks once a particular one is taken.

The subset of solutions we are interested in are those with
$\langle J^2 \rangle_0 \equiv J^2_0 > 0$, because as can be
seen from Eq.(\ref{evac}), in the presence of vacuum
fluctuations of physical currents, the nonlinear terms in
the equations of motion induce a nonzero stable
configuration $\langle \rho \rangle_0 \equiv v \neq 0$.
In this case we have
\begin{equation}
\label{vlu}
\mu^2\langle \rho \rangle_0\langle \rho^3 \rangle_0 =g^2
\langle J^2 \rangle_0
\end{equation}
On the other hand, Eq.(\ref{sol}) leads to
\begin{equation}\label{el2}
\langle J^2 \rangle_0=\langle \rho^4 \rangle_0\langle
A_\mu A^\mu \rangle_0
\end{equation}
so that, the solutions with $\langle \rho \rangle_0
\ne 0$ satisfy
\begin{equation}\label{laa}
\mu^2\langle \rho \rangle_0 \langle \rho^3 \rangle_0=
g^2\langle \rho^4 \rangle_0 \langle A_\mu A^\mu \rangle_0
\end{equation}
Notice that a similar equation, relating the
mass of the scalar field with the vacuum
fluctuations of the gauge field, can be directly obtained
from the initial equations of motion. Indeed, taking the
VEV of Eq.(\ref{ero}) one finds
\begin{equation}\label{como}
\mu^2\langle \rho \rangle_0 =
g^2\langle A_\mu A^\mu\rho\rangle_0
\end{equation}
However, physical considerations lead us to look for
solutions satisfying in addition
\begin{equation}\label{estas}
\langle A_\mu A^\mu\rho\rangle_0 \simeq
\langle A_\mu A^\mu\rangle_0 \langle\rho\rangle_0
\end{equation}
This condition, which obviously holds in a perturbative
phase, is necessary in order to allow us the construction
of a Hilbert-space basis with a simple
interpretation in terms of quanta of the $A^\mu(x)$ and
$\rho(x)$ fields, which eventually must be considered as
the physically relevant degrees of freedom.
Making use of this requirement in Eq.(\ref{como}), we
find that there exists physically meaningful solutions
with $\langle \rho \rangle_0 \neq 0$ only if
\begin{equation}\label{ule}
\langle A_\nu A^\nu \rangle_0 \simeq \mu^2/g^2
\end{equation}
This equation basically states that for the process to
take place, $\mu$ particles should be created from
energy fluctuations of gauge fields in the ground state,
and to the extent that $\mu^2$ is an externally given
parameter it may represent a strong constraint.

Note on the other hand that the existence of solutions
with a simple interpretation in terms of quanta of
$A^\mu(x)$ and $\rho(x)$ is not, in general,
compatible with Eq.(\ref{sol}). In fact, the existence
of such solutions is again a consequence of the rigidity
of the scalar field which, as will be seen, allows us to
absorb the nonperturbative part of the interaction into a
redefinition of the fields, leading to an effective
theory with a massive gauge field interacting with those
scalar particles situated above the ground state.
Comparing Eqs.(\ref{laa}) and (\ref{ule}) we find
that consistency demands
\begin{equation}
\langle \rho^4 \rangle_0 \simeq \langle \rho \rangle_0
\langle \rho^3 \rangle_0
\end{equation}
so that, we are led to look for solutions satisfying
\begin{equation}\label{vn}
\langle \rho^n \rangle_0 \simeq
\langle \rho \rangle_0^n = v^n
\end{equation}
which, as can be easily verified, represents a
sufficient condition in order for Eq.(\ref{estas}) to
hold in the nonperturbative phase described
by the solutions (\ref{sol}). This factorization
property, which in fact is also implicit in the Higgs
mechanism, is characteristic of a system with a
macroscopic occupation of the ground state, and
in particular implies that the behaviour of the scalar
field in the vacuum must be essentially classical
in character. Indeed, from Eq.(\ref{vn}) we have
\begin{equation}
\sqrt{\langle \rho^2 \rangle_0 -
\langle \rho \rangle_0^2} \simeq 0
\end{equation}
showing that our solutions are incompatible with the
existence of relevant scalar fluctuations regardless of
their thermal or quantum nature. This fact justifies a
zero-temperature treatment in determining the
asymmetric vacuum.

Making use of Eq.(\ref{vn}) we can rewrite
Eq.(\ref{vlu}) in the form
\begin{equation}\label{v4l2}
v^4={g^2 \over \mu^2} \langle J^2 \rangle_0
\end{equation}
which reflects that the appearance of a nonvanishing VEV
$\langle \rho \rangle_0$ is a direct consequence of the
existence of vacuum fluctuations of the physical current
$J^\mu (x)$.

Notice, on the other hand, that (\ref{vn}) also implies
that the vacuum constitutes a stable configuration of
$V_{\rm eff}(\rho ; J^2)$. More precisely,
from Eqs.(\ref{xw1}), (\ref{xw2}) and (\ref{vn}) one
finds that the VEVs of $\rho(x)$ and $J^2(x)$ minimize
the potential energy density
\begin{equation}\label{bubu}
\left.{{\partial \over \partial \rho} V_{\rm eff}}\right
\vert_{\langle { }
\rangle} \simeq 0
\end{equation}
where the notation $\langle { } \rangle$ stands for
$(\rho ,J^2)=(v, J^2_0)$.
In fact, as can be readily verified, Eq.(\ref{bubu}) is
nothing but a particular case of the more general relation
\begin{equation}\label{cond}
\left.{{\partial^n \over \partial \rho^n} V_{\rm eff}}\right
\vert_{\langle { }
\rangle} \simeq
\langle 0 \vert {\partial^n \over \partial \rho^n} V_{\rm eff}
\vert 0 \rangle
\end{equation}
which holds for $n \geq 0$.

As expected, expansion about the stable configuration
contributes to simplify considerably the theory.
By expanding $V_{\rm eff}(\rho ;J^2)$ about the vacuum
\begin{eqnarray}\label{vefant}
V_{\rm eff}(x) =&& \left. V_{\rm eff} \right
\vert_{\langle { } \rangle}
+\left. {\partial V_{\rm eff} \over \partial J^2}
\right \vert_{\langle { } \rangle} (J^2(x)-J^2_0)
+ \frac {1}{2!} \left. {\partial^2 V_{\rm eff}
\over \partial \rho^2} \right \vert_{\langle { }
\rangle} (\rho(x) - v)^2
\nonumber\\
&&+ \left. {\partial^2 V_{\rm eff} \over \partial \rho \:
\partial J^2} \right \vert_{\langle { } \rangle}
(\rho(x) - v) (J^2(x) - J^2_0)
+ \frac{1}{3!}
\left. {\partial^3 V_{\rm eff} \over \partial \rho^3}
\right \vert_{\langle { } \rangle} (\rho(x) - v)^3 +
\ldots
\end{eqnarray}
one obtains
\begin{eqnarray}\label{ant}
V_{\rm eff}(x) =&& \mu^2 v^2 + 2 \mu^2 (\rho(x) - v)^2
- {2 \mu^2 \over v} (\rho(x) - v)^3
+ {5 \mu^2 \over 2 v^2} (\rho(x) - v)^4
\nonumber\\
&& + {g^2 \over 2 v^2} (J^2(x) - J^2_0)
- {3 \mu^2 \over v^3} (\rho(x) - v)^5 -{g^2 \over v^3}
(\rho(x) - v)(J^2(x) - J^2_0) + O({1 \over v^4})
\end{eqnarray}
where use has been made of Eq.(\ref{v4l2}).
Higher terms are increasingly suppressed
by inverse powers of the VEV of the scalar field, so
that the physically relevant limit
$\langle \rho \rangle_0 \rightarrow \infty$
corresponds to a nonperturbative weak coupling phase.

Notice that up to corrections of order $1/v^2$ the
effect of gauge interactions is completely absorbed
into a redefinition of the low-energy effective scalar
theory. Indeed, substitution of Eq.(\ref{ant}) into
Eq.(\ref{rove}) leads to
%
\begin{equation}\label{mft}
\partial_\alpha \partial^\alpha \rho(x) + m^2 (\rho(x)-v)=
\lambda (\rho(x)-v)^2 + \ldots
\end{equation}
where
\begin{eqnarray}
m^2=&&\left.{\partial^2 \over \partial \rho^2} V_{\rm eff}
\right \vert_{\langle { } \rangle} =\mu^2 +
3 {g^2 l^2 \over v^4}=4\mu^2\label{masa}\\
\lambda=&& -\frac{1}{2}\left. {\partial^3 \over \partial
\rho^3} V_{\rm eff}
\right \vert_{\langle { } \rangle} =
6 {g^2 l^2 \over v^5}={6 \mu^2 \over v}
\end{eqnarray}
so that, as previously said, $\rho(x)$ decouples,
becoming unaffected by the presence of gauge (and matter
\cite{VDB1}) fields.
In fact Eq.(\ref{mft}) describes a low-energy mean field
theory involving an independent scalar field
$\rho(x)$ subject to an effective mean potential which
contains the effects of gauge interactions. In particular,
according to (\ref{cond}) the mass and coupling constant
characterizing the dynamical evolution of $\rho(x)$ are
simply given by
\begin{eqnarray}
m^2 \simeq &&\langle 0 \vert {\partial^2 \over \partial
\rho^2} V_{\rm eff} \vert 0 \rangle \\
\lambda \simeq && -\frac{1}{2}\langle 0 \vert {\partial^3
\over \partial \rho^3} V_{\rm eff} \vert 0 \rangle
\end{eqnarray}

Experience tells us that mean field theories usually
describe weakly interacting systems with a long-range
order characteristic of macroscopic classical systems.
Not surprisingly the low-energy scalar theory defined
by Eq.(\ref{mft}) exhibits long-range order too. Indeed,
the factorization property (\ref{vn}) implies that the
correlation function $\langle 0 \vert \rho(x) \rho(y)
\vert 0 \rangle$ remains constant over arbitrary large
space-time intervals
\begin{equation}
\langle 0 \vert \rho(x) \rho(y) \vert 0 \rangle
\simeq v^2
\end{equation}
Such infinite-range behaviour is a characteristic
feature of superfluid Bose systems \cite{Pines} and is
a consequence of the macroscopic occupation of a single
quantum state.
As we shall see below, the scalar field $\rho(x)$
may be interpreted as describing an almost uniform
superfluid composed of $\mu$ particles mainly
condensed in the zero-momentum state, which represents
the lowest-energy state and therefore defines the
nontrivial vacuum of the theory.

As Eq.(\ref{mft}) shows, in a dynamical description
rather than $\rho(x)$ itself the physically relevant
quantity is its departure from the vacuum. It is
therefore convenient to define a real scalar field with
vanishing VEV
\begin{equation}\label{dfeta}
\eta(x)=\rho(x)-v
\end{equation}
in terms of which the equations of motion governing the
low-energy effective theory finally read
\begin{eqnarray}
\label{effr}
&&(\partial_\alpha \partial^\alpha  + m^2) \eta(x)= \lambda
\eta^2(x)
+ \ldots \\
\label{eaf}
&&(\partial_\alpha \partial^\alpha + m^2_A )A^\nu -\partial^\nu
(\partial_\beta A^\beta) = j^\nu(x)
\end{eqnarray}
where
\begin{equation}\label{jcor}
j^\nu(x) = -(2g m_A \eta(x)+g^2\eta^2(x)) A^\nu(x)
\end{equation}
Therefore for a particular solution (\ref{sol}) with
$J^2_0 > 0$, the scalar field develops a nonvanishing
VEV
\begin{equation}
\phi(x)= {1\over \sqrt {2}}(v+\eta(x)) \exp i\xi(x)
\end{equation}
and the gauge field $A_\mu(x)$ acquires a mass
\begin{equation}
\label{mbos}
m^2_A=g^2 v^2
\end{equation}
in accordance with the well-known results \cite{libro}.
Nonzero vacuum fluctuations of $J^\mu(x)$ induce
spontaneous breakdown of gauge symmetry, which
then becomes hidden in the sense that the particle
spectrum of the low-energy effective theory does not
display the full symmetry of the original Lagrangian
density.

Eq.(\ref{effr}) reflects the fact that the field
$\rho(x)=v+\eta(x)$ decouples.
Quantum fields exhibit this kind of behaviour in the
limiting case where a macroscopically large number of
quanta appear in a coherent state, and in such a
situation they behave as ordinary classical fields.
In particular, $\rho(x)$, which describes $\mu$
particles, behaves as a classical scalar field that
fluctuates about $\rho(x)=v$ with a mass $m=2\mu$. Then
one expects that a macroscopic physical meaning should be
possible to be given to $\rho(x)$.
Indeed, it may be interpreted as describing a superfluid
Bose condensate of $\mu$ particles with vanishing momenta
and a particle number density
\begin{equation}
\label{densi}
n(x)=\frac {1}{2} m \rho^2(x)
\end{equation}
In order to see this, let us consider a real scalar field
$\rho(x)$ describing a system of almost non-interacting
bosons of mass $m$ in a volume $V \rightarrow \infty$.
At zero temperature most particles are condensed in
the zero-momentum state, which is the lowest-energy
state. Under this conditions, because of the fact that
in replacing
\begin{equation}
{1 \over V} \sum_{\bf k} \rightarrow {1 \over
(2 \pi)^3} \int {d^3 {\bf k}}
\end{equation}
one neglects the contribution of the zero-momentum
state, a continuous treatment turns out to be
inadequate. Therefore we shall use in what follows a
discrete formulation, which proves to be particularly
convenient in treating highly degenerate Bose systems.

In the case of almost non-interacting particles,
$\rho(x)$ may be expanded in a Fourier series of plane
waves in the form
\begin{equation}\label{teide}
\rho(x)=\sum_{\bf k}{1 \over \sqrt{V \, 2k^0}}
[a_{\bf k} e^{-ikx} + a^{+}_{\bf k} e^{ikx}]
\end{equation}
where $k^0=(m^2 + {\bf k}^2)^{1/2}$, $kx=k^0 x^0
-{\bf kx}$, and $a^{+}_{\bf k}$, $a_{\bf k}$ are the
creation and annihilation operators satisfying the
usual conmutation relations
\begin{equation}
[a_{\bf k},a^{+}_{{\bf k} \prime} ]= \delta_{{\bf k}
{\bf k}\prime}
\end{equation}
The state of the system, $\vert \alpha \rangle$, may be
completely characterized by giving the occupation numbers,
$N_{\bf k}$, of the different {\bf k}-momentum states.
In particular, the vacuum $\vert 0 \rangle$ is simply
defined by the condition $N_{\bf k} = 0$ for all
${\bf k} \neq {\bf 0}$
(i.e., no particles in excited states)
\begin{equation}
\vert 0 \rangle \equiv \vert N_0,0,\ldots,0,\ldots
\rangle
\end{equation}
Since at temperature $T \simeq 0$ most particles
are condensed in the zero-momentum state, we have
\begin{equation}\label{ifara}
N_0 \gg N_{{\bf k}\ne {\bf 0}}
\end{equation}
On the other hand, because of the macroscopic size of
the system ($V \rightarrow \infty$) we expect in
addition a macroscopic occupation of the lower excited
states
\begin{equation}
N_0,\ldots, N_{\bf k}, \ldots \gg 1
\end{equation}
Under these conditions, as first noted by Bogoliubov
\cite{Bogo}, the corresponding creation and
annihilation operators, $a_{\bf k}^+$ and $a_{\bf k}$,
behave as c-numbers. Indeed, for a macroscopically
occupied {\bf k}-momentum state the relations
\begin{eqnarray}
a_{\bf k} \, \vert N_0,\ldots,N_{\bf k},\ldots \rangle=&&
\sqrt{N_{\bf k}} \,\, \vert N_0 \ldots,N_{\bf k}-1,\ldots
\rangle
\\
a_{\bf k}^+ \, \vert N_0,\ldots,N_{\bf k},\ldots \rangle=&&
\sqrt{N_{\bf k}+1} \,\, \vert N_0 \ldots,N_{\bf k}+1,\ldots
\rangle
\end{eqnarray}
simply reduce to
\begin{eqnarray}
a_{\bf k} \, \vert \alpha \rangle \simeq &&
\sqrt{N_{\bf k}} \, \vert \alpha \rangle
\\
a_{\bf k}^+ \, \vert \alpha \rangle \simeq &&
\sqrt{N_{\bf k}} \, \vert \alpha \rangle
\end{eqnarray}
where we have made use of the fact that the state of
the system
\begin{equation}
\vert \alpha \rangle \equiv \vert N_0,\ldots, N_{\bf k},
\ldots \rangle
\end{equation}
remains virtually unaffected by the addition or
substraction of a particle in the {\bf k}-momentum state
(the corresponding correction is only of order
$1/N_{\bf k}$).
Therefore, the operators $a_{\bf k}^+$ and $a_{\bf k}$
behave as c-numbers and may be replaced by
$\sqrt{N_{\bf k}}$.
Using this approximation in Eq.(\ref{teide}) we obtain
\begin{equation}\label{anaga}
\rho(x)=\sqrt{{2n_0\over m}} \, \cos(mt) +
\sum_{{\bf k}\ne {\bf 0}} {1 \over \sqrt{V \, 2k^0}}
[a_{\bf k} e^{-ikx} + a^{+}_{\bf k} e^{ikx}]
\end{equation}
where $n_0$ denotes the mean number density of the ground
state
\begin{equation}
n_0 \equiv {N_0 \over V}
\end{equation}
Eq.(\ref{anaga}) may be written in the form
\begin{equation}\label{masca}
\rho(x)=\sqrt{{2n_0\over m}} \, \bigl[ 1+O(m^2 t^2) \bigr]
+ \eta(x)
\end{equation}
where the classical field $\eta(x)$ is given by
\begin{equation}\label{adeje}
\eta(x)=\sum_{{\bf k}\ne {\bf 0}} {1 \over
\sqrt{V \, 2k^0}}
[a_{\bf k} e^{-ikx} + a^{+}_{\bf k} e^{ikx}]
\simeq \sum_{{\bf k}\ne {\bf 0}} \sqrt{{2n_k\over k^0}}
\, \cos(k^0 x^0-{\bf kx})
=\langle \alpha \vert \eta(x) \vert \alpha \rangle
\end{equation}
$n_k$ being the mean number density of the
{\bf k}-momentum state. Incidentally, note that
$\langle 0 \vert \eta(x) \vert 0 \rangle = 0$.

Comparing (\ref{dfeta}) with (\ref{masca}) in the
limit $mt \ll 1$ \cite{nota}, one is led to identify
\begin{equation}\label{tinguaro}
v=\sqrt{{2n_0\over m}}
\end{equation}
which provides a physical interpretation for
$\langle \rho \rangle_0$. Indeed, this equation
states that the VEV of $\rho(x)$ essentially measures the
mean number density of particles condensed in the
zero-momentum state (vacuum).
Eq.(\ref{tinguaro}) also reflects that the limit
$v \rightarrow \infty$ physically corresponds to a
macroscopically occupied ground state, $n_0 \rightarrow
\infty$. Therefore the factorization property (\ref{vn}),
which holds provided that
\begin{equation}
\langle 0 \vert \eta^n(x) \vert 0 \rangle \ll v^n
\end{equation}
is indeed a consequence of the macroscopic occupation of
the vacuum. Making use of this property we may rewrite
Eq.(\ref{tinguaro}) in the form
\begin{equation}
n_0 \simeq \langle 0\vert\frac{1}{2}m\rho^2(x)\vert 0
\rangle
\end{equation}
so that one expects the number density of the
Bose system in the state $\vert \alpha \rangle$ to be
given by
\begin{equation}\label{ofra}
n(x) \equiv \langle \alpha \vert\frac{1}{2}m\rho^2(x)
\vert \alpha \rangle \simeq \frac{1}{2}m\rho^2(x)
\end{equation}
where the last step reflects the fact that, to a good
approximation, $\rho(x)$ behaves as a c-number.
Taking into account that
$\langle \alpha \vert\eta(x)\vert \alpha \rangle \sim
\sum \sqrt{n_k}$ while $v \sim \sqrt{n_0}$, one has,
according to Eq.(\ref{ifara}),
\begin{equation}
\langle \alpha \vert\eta(x)\vert \alpha \rangle \ll v
\end{equation}
so that (\ref{ofra}) can be expressed in the form
\begin{equation}\label{fluc}
n(x) \simeq n_0 \left[ 1 + 2 {\eta(x) \over v} \right]
\end{equation}
with $\eta(x)/v \ll 1$.
Then, the scalar background field $\rho(x)=v+\eta(x)$
may be interpreted as describing an almost uniform
(i.e., homogeneous and isotropic) superfluid condensate
of decoupled $\mu$ particles with a number density $n(x)
\simeq n_0$, the superfluid behaviour being a consequence
of the fact that most particles are condensed in the
zero-momentum state and therefore do not contribute to
the viscosity (neither to the pressure).

Eq.(\ref{fluc}) also leads to
\begin{equation}
\eta(x) \simeq {v \over 2n_0} (n(x)-n_0)
\end{equation}
which shows that $\eta(x)$ can be regarded in turn as
describing density fluctuations about the mean value
$n_0$. Incidentally, note that according to
Eq.(\ref{masa}) the mass of this field corresponds to
the energy necessary to create a pair of $\mu$ particles.

With the present interpretation the minimun of the
potential energy density may be written as
\begin{equation}\label{oye}
\left. V_{\rm eff} \right \vert_{\langle { } \rangle}=
\mu^2 v^2= \mu n_0
\end{equation}
where $n_0$ is the mean number density of the ground
state. Therefore,
$\left. V_{\rm eff} \right \vert_{\langle { } \rangle}
\simeq \langle 0 \vert V_{\rm eff} \vert 0 \rangle$
[see Eq.(\ref{cond})] is nothing but the internal energy
of a system of non-interacting $\mu$ particles condensed
in the zero-momentum state.

On the other hand, according to Eqs.(\ref{mbos}) and
(\ref{tinguaro}), the gauge boson acquires a mass
\begin{equation}
m_A^2 = {g^2 n_0 \over \mu}
\end{equation}
which also agrees with what one would expect.
Indeed $m_A^{-1}$ coincides with the London penetration
depth for a gauge field in a condensate of $\mu$
particles with number density $n_0$. Therefore, the
physical parameter $m_A^2$ accounts for the cumulative
effect of nonperturbative interactions with those scalar
particles condensed in the zero-momentum state, and
the low-energy ($T \simeq 0$) effective theory governed
by the equations of motion (\ref{effr})--(\ref{eaf})
describes a gauge field which evolves in a classical scalar
medium, acquiring mass and producing in turn density
fluctuations in this medium.
This is in fact a natural mechanism for particles to get
mass. Indeed experience shows that particles in dense
media respond with a larger inertia.

In the broken phase, the effective Lagrangian density
governing the dynamics of the quantum field $A_\mu(x)$
in the presence of the scalar background field
$\rho(x)=v+\eta(x)$ takes the form
\begin{equation}\label{Lq}
{\cal L}_{\rm eff}^{q} = - \frac{1}{4} {\rm F}_{\mu\nu}
{\rm F}^{\mu\nu} +
\frac{1}{2} m_A^2 A_\mu A^\mu + {\cal {L}_{\rm I}}
\end{equation}
where
\begin{equation}
{\cal {L}_{\rm I}} = gm_A\eta A_\mu A^\mu
+\frac {1}{2} g^2 \eta^2 A_\mu A^\mu
\end{equation}
describes the interaction of $A_\mu(x)$ with the external
classical field $\eta(x)$.
In turn, the decoupled evolution of the hidden scalar
medium becomes governed by the effective Lagrangian
density
%
\begin{equation}\label{Leff}
{\cal L}_{\rm eff}^{c}=\frac {1}{2} (\partial_\mu \eta)
(\partial^\mu \eta)
-\left(\left. V_{\rm eff}\right\vert_{\langle { }\rangle}+
\frac{1}{2} m^2 \eta^2 - {\lambda \over 3} \eta^3 +
\ldots \right)
\end{equation}
which, as previously stated, defines a mean field theory
containing the effects of gauge interactions.

The decoupling of the scalar sector enables
us to define the energy density of the medium
\begin{equation}
{\cal H}_{\rm eff}^c =\frac{1}{2} {\dot{\eta}}^2 +
\frac{1}{2} (\nabla \eta)^2 + V_{\rm eff}(x)
\end{equation}
so that, according to Eq.(\ref{oye}), the vacuum
energy density turns out to be
\begin{equation}
\langle 0 \vert {\cal H}_{\rm eff}^c \vert 0 \rangle
\simeq \mu n_0
\end{equation}
indicating that, to a good approximation, the chemical
potential of the Bose superfluid at $T \simeq 0$ coincides
with the mass $\mu$.

{}From Eqs.(\ref{Lq})--(\ref{Leff}) we see that
nonperturbative interactions with those
particles condensed in the zero-momentum state have been
absorbed into the physical parameters characterizing the
low-energy effective theory, ${\cal L}_{\rm eff}^q \oplus
{\cal L}_{\rm eff}^c$, which now describes elementary
excitations (particles) with respect to the nontrivial
asymmetric vacuum (ether). As a result, the vacuum degrees
of freedom disappear from the formulation.
The physical reason for this is related with the
nonperturbative decoupling of the scalar sector.
Indeed, according to Eqs.(\ref{ant}) and (\ref{tinguaro}),
in the broken phase the effective couplings become
suppressed by inverse powers of the number density $n_0$
of particles condensed in the ground state, so that
in the presence of a macroscopically large number
of quanta, $n_0 \rightarrow \infty$, the scalar field
decouples and then behaves as a classical field, just in
agreement with physical expectations. Under these conditions
the condensate appears as a {\em rigid} (i.e.,
unperturbed) macroscopic classical body, so that, as
usually happens in these cases, its effects only enters
the formulation in a static way.

On the other hand, as evident from Eq.(\ref{ant}), in the
limit $n_0 \rightarrow \infty$ the scalar sector not only
decouples but also becomes non-interacting ({\em
trivial}\,). It defines a nonperturbative low-temperature
free scalar theory.
However, due to the fact that in the present mechanism the
symmetry breakdown occurs as a direct consequence of gauge
interactions (rather than as a consequence of
self-interactions in the scalar sector) obviously the
{\em triviality} of the theory poses no problem here.

It should be emphasized that the low-energy effective
theory ${\cal L}_{\rm eff}^q \oplus {\cal L}_{\rm eff}^c$
is only adequate for describing the broken phase
characterized by the nonvanishing order parameter
$\langle \rho \rangle_0 \equiv v \ne 0$.
In fact, both Eq.(\ref{evac}) and the expansion
(\ref{ant}) are nonanalytic in the point $v=0$,
indicating the nonperturbative character of our
treatment. The physical reason for this behaviour is
quite clear.
Since the vanishing of the order parameter corresponds
to the restoration of gauge symmetry, the limit
$v \rightarrow 0$ represents a phase transition from
the broken phase to the symmetric one. However,
experience shows that phase transitions are in general
quite complex physical phenomena where the physical
properties of the system usually change drastically.
Under these conditions it is not possible to continously
pass from one to another phase and a specific
mathematical formulation is required for describing the
asymmetric phase.
In fact the appearance of a nonvanishing order parameter
determines the emergence of new physics at the
low-temperature scale. A stable, highly-ordered
macroscopic medium with a clear classical behaviour
appears as an {\em emergent property} (i.e., a
property of a complex system which is not contained in
its parts \cite{Ander1}).

It should be noticed, however, that a proper treatment
of the phase transition leading to the restoration of
gauge symmetry would require a finite-temperature
formalism.
The possibility of phase transitions in gauge theories
was first suggested by Kirzhnitz \cite{Kir} and is
treated in detail, within the framework of the Higgs
mechanism, in Refs.\cite{PTI,PT1,PT2,PTF}.

To sum up,
starting from a gauge-invariant Lagrangian density
${\cal L}$ we have obtained, as a particular solution
of the equations of motion, a low-energy effective
theory, ${\cal L}_{\rm eff}^q \oplus
{\cal L}_{\rm eff}^c$, where ${\cal L}_{\rm eff}^q$
describes massive gauge bosons in a scalar medium and
would be relevant to particle physics, while
${\cal L}_{\rm eff}^c$ governs the decoupled dynamics
of the medium and would be relevant to cosmology.

-

I thank P. M. Stevenson for interesting remarks concerning
the $\phi^4$ interaction. I am also grateful to C. Mu\~noz
for valuable comments and support.


\begin{references}
\bibitem{nota}
In the phenomenologically relevant case of a
SU(3)$\times$SU(2)$\times$U(1) gauge theory it is
possible to  break the electroweak symmetry starting
with a massless scalar field, and the Higgs boson turns
out to be the pseudo-Nambu-Goldstone boson responsible
for breaking the scale invariance of the low-energy
effective scalar theory. Therefore, mass appears as
a general dynamical property.
See Ref.\cite{VDB1}.
\end{references}
\end{document}